\documentclass[preprint2]{emulateapj}
\usepackage{natbib}
\usepackage{amsmath}    % need for subequations



\bibpunct[; ]{(}{)}{,}{a}{}{;}

\shorttitle{Continuum Contributions to the SDO/AIA Passbands During Solar Flares}

\shortauthors{Milligan \& McElroy}

\begin{document}

\title{Continuum Contributions to the SDO/AIA Passbands During Solar Flares}
\notetoeditor{the contact email is r.milligan@qub.ac.uk, is the only one which should appear on the journal version}
\author{Ryan O. Milligan\altaffilmark{1,2,3} \& Sarah A. McElroy\altaffilmark{1}}

\altaffiltext{1}{Astrophysics Research Centre, School of Mathematics \& Physics, Queen's University Belfast, University Road, Belfast, Northern Ireland, BT7 1NN}
\altaffiltext{2}{Department of Physics, Catholic University of America, 620 Michigan Ave., N.E., Washington, DC 20064 }
\altaffiltext{3}{Solar Physics Laboratory (Code 671), Heliophysics Science Division, NASA Goddard Space Flight Center, Greenbelt, MD 20771, USA}

\begin{abstract}
\noindent
Data from the Multiple EUV Grating Spectrograph (MEGS-A) component of the Extreme Ultraviolet Experiment (EVE) onboard the Solar Dynamics Observatory (SDO) were used to quantify the contribution of continuum emission to each of the EUV channels of the Atmospheric Imaging Assembly (AIA), also on SDO, during an X-class solar flare that occurred on 2011 February 15. Both the pre-flare subtracted EVE spectra and fits to the associated free-free continuum were convolved with the AIA response functions of the seven EUV passbands at 10 s cadence throughout the course of the flare. It was found that 10--25\% of the total emission in the 94\AA, 131\AA, 193\AA, and 335\AA\ passbands throughout the main phase of the flare was due to free-free emission. Reliable measurements could not be made for the 171\AA\ channel due to possible coronal dimming during the event, while the continuum contribution to the 304\AA\ channel was negligible due to the presence of the strong \ion{He}{2} emission line. Up to 50\% of the emission in the 211\AA\ channel was found to be due to free-free emission around the peak of the flare, while an additional 20\% was due to the recombination continuum of \ion{He}{2}. The analysis was extended to a number of M- and X-class flares and it was found that the level of free-free emission contributing to the 211\AA\ passband increased with increasing GOES class. These results suggest that the amount of continuum emission that contributes to AIA observations during flares is more significant than that stated in previous studies which used synthetic, rather than observed, spectra. These findings highlight the importance of spectroscopic observations carried out in conjunction with those from imaging instruments so that the data are interpreted correctly.
\end{abstract}


\section{Introduction}
\label{intro}
Since its launch in February 2010, the Atmospheric Imaging Assembly (AIA; \citealt{leme12}) onboard the Solar Dynamics Observatory (SDO; \citealt{pesn12}) has been producing high-resolution, full-disk images of the Sun in ten wavelength bands with a 12 second cadence. This instrument has a similar spatial resolution to that of former solar imager, the Transition Region and Coronal Explorer (TRACE; \citealt{hand99}), but has an increased field of view, temporal resolution and spectral coverage. Seven of the ten passbands are tuned to observe emission in the extreme ultraviolet (EUV; 94\AA, 131\AA, 171\AA, 193\AA, 211\AA, 304\AA, and 335\AA) and each of these passbands contain emission lines formed at distinctly different temperatures, as well as potential continuum emission. Knowledge of which emission processes dominate the images under different conditions are crucial for interpreting the data correctly.

Analysis of the emission contributing to the 171\AA\ and 195\AA\ channels on TRACE during flares was previously carried out by \cite{phil05} using the CHIANTI atomic database (Version 5; \citealt{dere97}). Typically, these channels are dominated by emission at quiescent coronal temperatures (1--2~MK). During flares, however, the 195\AA\ channel becomes sensitive to emission from the \ion{Fe}{24} line at 192.03\AA\ which is formed at $\sim$15~MK. Above these temperatures continuum emission was found to dominate the 171\AA\ channel. In the 195\AA\ channel the maximum contribution from continuum emission was estimated to be about 10\% at a temperature of 4~MK. \cite{feld99} and \cite{warr01} also showed that the continuum contribution to the 171\AA\ channel introduced further ambiguities when determining flare temperatures using filter ratio techniques.

Theoretical analysis of the composition of the SDO/AIA images has been conducted by \cite{odwy10}. The authors examined the contribution of both line and continuum (free-free) emission to each of the EUV channels for quiet Sun, coronal hole, active region and flare plasmas by convolving synthetic spectra, obtained using the CHIANTI atomic database (Version 6.0.1; \citealt{dere09}) under an assumed differential emission measure (DEM), with the response function for each AIA channel \citep{boer12}. In the case of flaring plasma, the DEM of \cite{dere79} from the decay phase of an M2 flare was used. Coronal abundances from \cite{feld92}, ionization equilibrium, and an electron density of 10$^{11}$~cm$^{-3}$ were also assumed. Under these conditions the authors found that the continuum contribution to each channel was negligible ($<$6\%), except in the case of the 171\AA\ and 211\AA\ channels were it was found be as high as 23\% and 41\%, respectively.

\begin{figure*}[!t]
	\begin{center}
	\includegraphics[width=\textwidth]{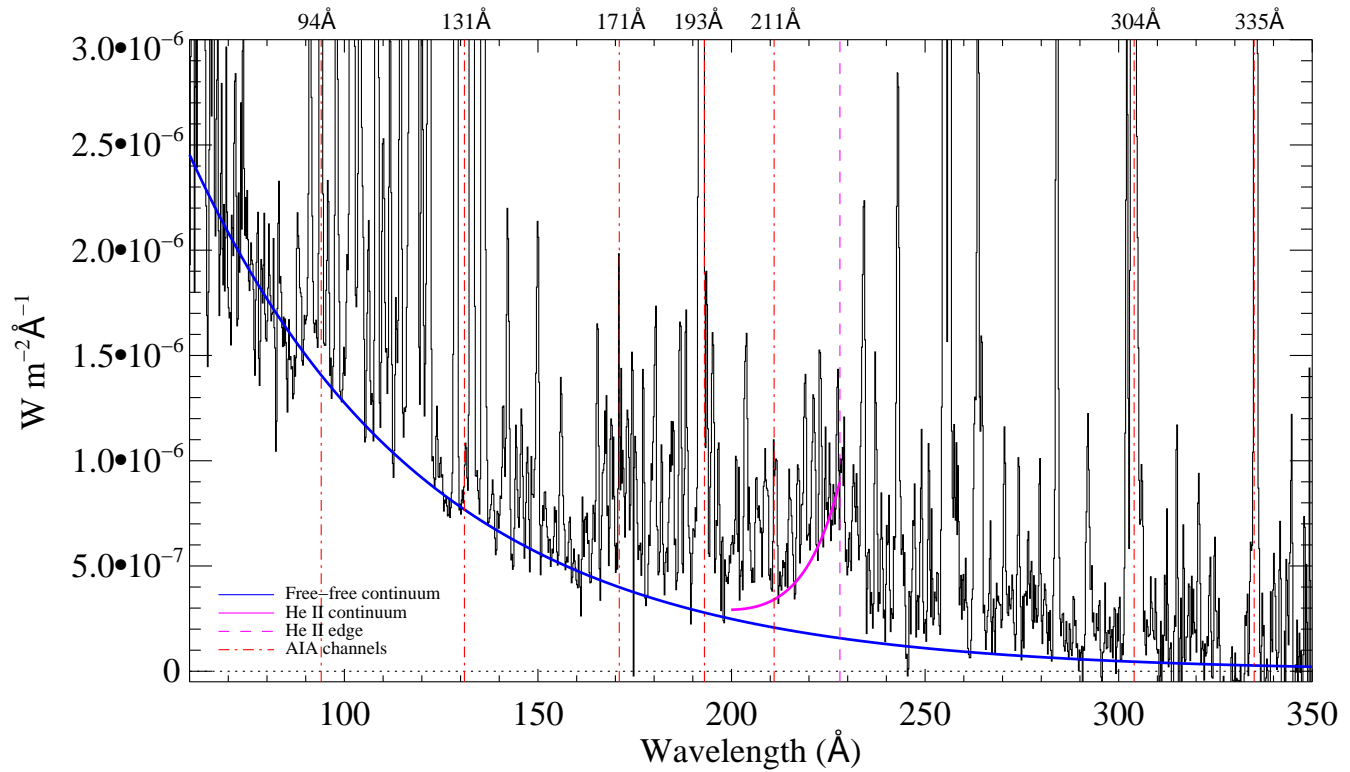}
	\caption{Background-subtracted EVE spectrum over the 60\AA-350\AA\ wavelength range taken near the SXR peak of the X-class flare on 2011 February 15. Overlaid in blue is the corresponding fit to the free-free continuum, while the purple curve denotes the fit to the \ion{He}{2} continuum blueward of the recombination edge at 228\AA\ (vertical dashed purple line). The vertical red dot-dashed lines are positioned at the central wavelength of each AIA passband.}
	\label{eve_spec}
	\end{center}
\end{figure*}

More recently, however, \cite{mill12} used spectral data from the EUV Variability Experiment (EVE; \citealt{wood12}), also on SDO, to show that continuum emission, both free-free (due to thermal bremsstrahlung) and free-bound (due to recombination), became significantly enhanced during an X-class solar flare. It was shown that the lightcurves of the free-free emission rose in concert with that of the GOES 1--8\AA\ lightcurve suggesting that the emission is predominantly coronal in nature, while the free-bound emission from \ion{H}{0} (Lyman continuum) and \ion{He}{1} peaked around the time of HXR emission as observed by the Ramaty High-Energy Solar Spectroscopic Imager (RHESSI; \citealt{lin02}) indicating a chromospheric origin. While emission from the \ion{He}{2} continuum was also believed to be chromospheric, it was found to be inherently weaker than the Lyman and \ion{He}{1} continua, and its lightcurve was found to peak much later in the flare due to the dominance of the underlying free-free emission during the flare's early stages. 

As the EVE MEGS-A (Multiple EUV Grating Spectrographs) wavelength range (60--370~\AA) contains each of AIA's EUV passbands (vertical dot-dashed red lines in Figure~\ref{eve_spec}) it is now possible to quantify the continuum contributions to each channel during solar flares by convolving the observed EVE spectra, rather than a synthetic one, with the AIA response functions. This not only removes any assumptions about the physical conditions of the emitting plasma, but also allows the relative continuum contribution to be determined as a function of time throughout a flaring event, rather than just during the decay phase as necessitated by the use of the flare DEM of \cite{dere79} by \cite{odwy10}. 

This paper presents a detailed analysis of the continuum contributions to each of the EUV channels on AIA during the X-class flare that occurred on 2011 February 15, as well as extending the analysis to a range of other large flares. Section~\ref{data_anal} describes the flare observations from EVE and the AIA response functions. The results are presented in Section~\ref{results}, and summarized in Section~\ref{conc} where their implications are also discussed.

\section{Observations and Data Analysis}
\label{data_anal}

The EVE instrument acquires full disk (Sun-as-a-star) EUV spectra every 10 seconds over the 65--370~\AA\ wavelength range using its MEGS-A component with a near 100\% duty cycle. In March 2013, Version 3 of the Level 2 data (used in this work) were released. \cite{delz13} noted that irradiance values around 100\AA\ in Version 3 are $\sim$50\% lower than those in Version 2, which were used by \cite{mill12}.

\begin{figure*}[!t]
	\begin{center}
	\includegraphics[width=\textwidth]{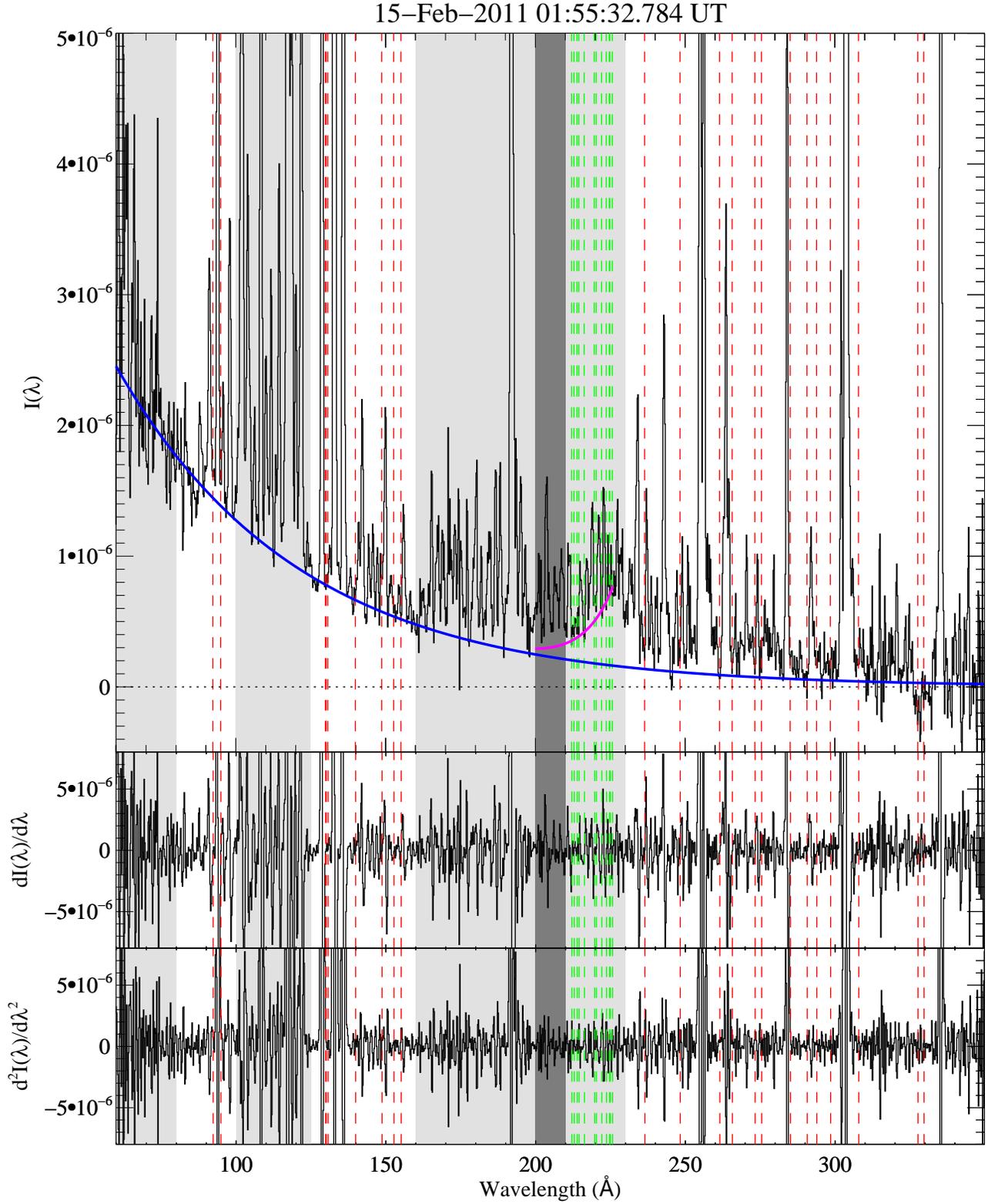}
	\caption{Example of how EVE spectra were fit in order to quantify the continuum emission. Top panel: pre-flare subtracted EVE MEGS-A spectrum (I($\lambda$)) from 60--350\AA\ from the SXR peak of the 2011 February 15 flare. Bottom two panels: first and second derivatives of the spectrum with respect to wavelength. The vertical dashed red lines denote the local minima of I($\lambda$), where $dI(\lambda)/d\lambda \approx 0$ and $d^{2}I(\lambda)/d\lambda^{2}>0$. These data points were fit with an exponential function with which to represent the free-free continuum (solid blue curve in top panel). Similarly, the vertical dashed green lines denote the local minima over the 210--228\AA\ range (having subtracted the fit to the free-free emission) with which to quantify the \ion{He}{2} continuum (purple curve). The grey shaded areas illustrate the portions of the spectra which were omitted during the fitting process (see text).}
	\label{deriv_spec}
	\end{center}
\end{figure*}

On 2011 February 15 the first X-class flare of Solar Cycle 24 occurred; an X2.2 flare that began at 01:44~UT (SOL2011-02-15T01:56). Figure~\ref{eve_spec} shows an EVE MEGS-A spectra (pre-flare subtracted) taken near the SXR peak of the flare (01:55:32 UT). The pre-flare irradiance was calculated by averaging over 90~s prior to the flare onset (from 01:00:12--01:01:42~UT) and was subtracted from the total irradiance to isolate the flux produced by the flare itself. This figure shows that the underlying free-free emission (represented by the solid blue curve) spans the entire MEGS-A spectral range, and that the enhancement due to the flare is greater at shorter wavelengths. The recombination continuum of \ion{He}{2} (purple curve blueward of the recombination edge at 228\AA; vertical dashed purple line) is also visible and appears to straddle the 211\AA\ passband.

Measurements of the continuum emission during this flare were previously deduced by \cite{mill12} by fitting the line-free portions of the EVE MEGS-A spectrum (deduced from a synthetic flare spectrum from CHIANTI) with an exponential function. However, given that the EVE wavelength range is likely to contain emission lines which are not present in the CHIANTI database ({\it c.f.} \citealt{brow08}) it is possible that a number of weak lines may have biased the previous fits to the continuum. Furthermore, this technique was found to produce inconsistant continuum time profiles for other events. And so rather than assuming that a limited number of spectral ranges were free from any emission lines, it was assumed that the free-free emission corresponded to the lower envelope of the EVE spectra. Therefore the local minima ($\lambda_{i}$) across the MEGS-A spectral range were identified by taking the first and second derivatives of the EVE spectra ($I(\lambda,t)$; bottom two panels of Figure~\ref{deriv_spec}) with respect to wavelength, such that:
\begin{equation*}
\frac{dI(\lambda,t)}{d\lambda} \approx 0 \hspace{0.5cm} \mbox{and} \hspace{0.5cm} \frac{d^2I(\lambda,t)}{d\lambda^2} > 0,
\end{equation*}
and 
\begin{equation*}
\frac{dI(\lambda_{i+1},t)}{d\lambda} > \frac{dI(\lambda_{i-1},t)}{d\lambda},
\end{equation*}
where $i$ is the index of the wavelength bin at which the first two conditions were met.

These local minima (vertical dashed red lines in Figure~\ref{deriv_spec}) were then fit with an exponential function (solid blue curve) similar to that of \cite{mill12}. However, three wavelength ranges (light shaded areas in Figure~\ref{deriv_spec}) were omitted from the fitting process in order to prevent possible biases: 60\AA$<\lambda<$80\AA\ was omitted due to scatter in the background-subtracted data at these wavelengths; 100\AA$<\lambda<$125\AA\ was omitted to avoid the pseudo-continuum generated by the close proximity of \ion{Fe}{19}--\ion{Fe}{23} lines; and 160\AA$<\lambda<$230\AA\ was omitted to avoid any negative flux values around 171\AA\ due to possible coronal dimming \citep{wood11}, and free-bound emission blueward of the \ion{He}{2} edge at 228\AA. 

Having fit the free-free continuum, this was then subtracted from the data to allow the fitting of the free-bound continuum of \ion{He}{2} (assumed to extend from 200-228\AA) using a similar process. The 200--210\AA\ range (dark shaded region in Figure~\ref{deriv_spec}) was omitted from the fitting process due to an elevated pseudo-continuum, while the local minima between 210-228\AA\ (vertical dashed green lines in Figure~\ref{deriv_spec}) were fit with an exponential function. However, these fits were only carried out up until the end time of the flare as stipulated in the GOES event list\footnotemark[1]\footnotetext[1]{The end time of a GOES event is defined as the time when the flux level in the 1--8\AA\ channel decays to a point halfway between the maximum flux and the pre-flare background level.}, beyond which the data became too noisy to reliably measure any recombination emission.

Both the pre-flare subtracted EVE data and the fits to the underlying continua were convolved with the AIA response functions at each 10~s interval throughout the course of the flare to yield a measurement of the contributions to each channel. The continuum was subtracted from the total irradiance in each channel to establish the relative contributions from both line and continuum emission.

\begin{figure*}[!t]
	\begin{center}
		\includegraphics[width=18.5cm]{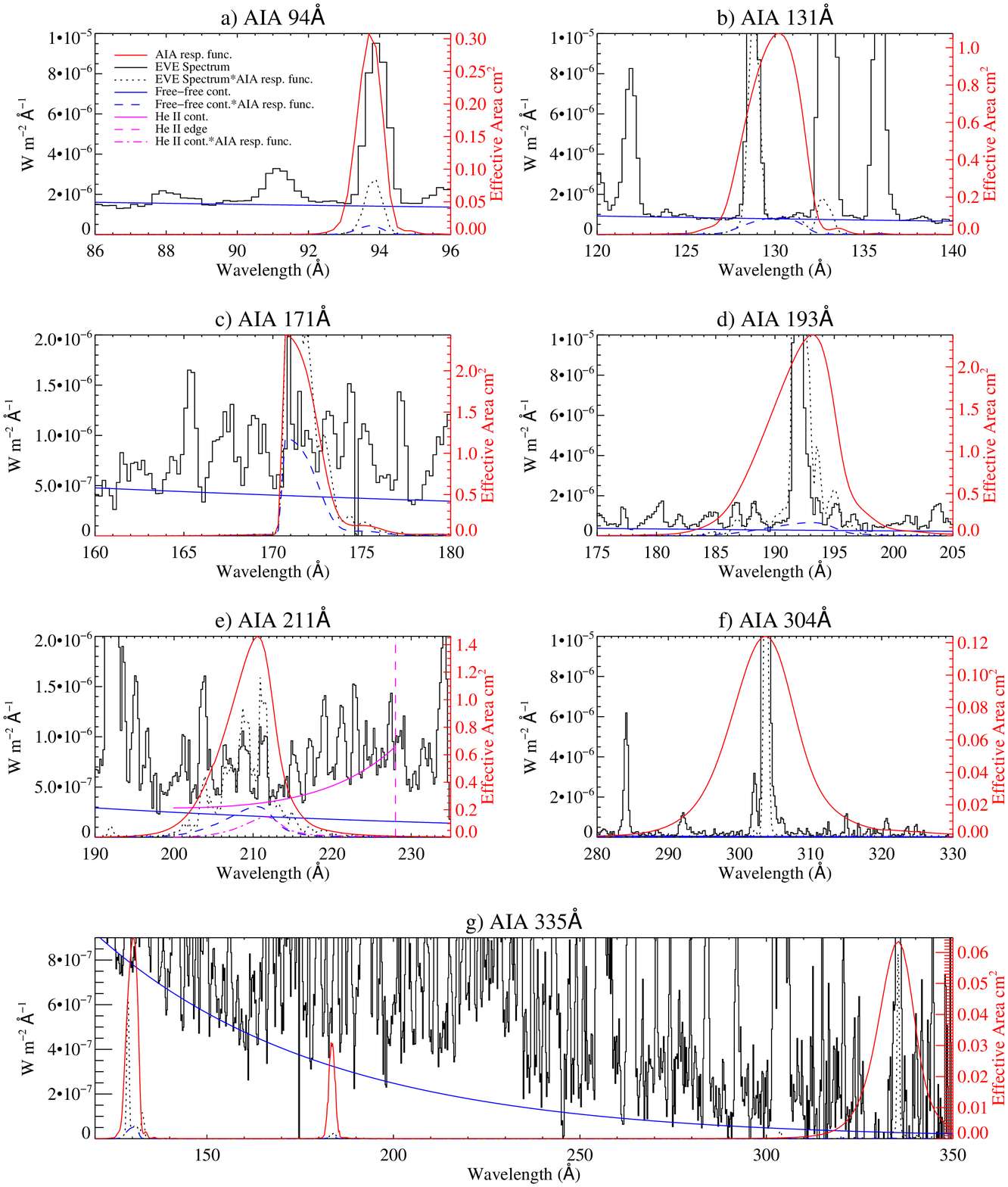}
		\caption{Plots of the EVE spectra (black histograms) and fits to the free-free continuum (solid blue line) around each of the seven EUV passbands on AIA at the SXR peak of the 2011 February 15 flare. In panel $e$ the fit to the \ion{He}{2} continuum is also shown. The AIA effective areas are overplotted as solid red lines. The convolved profiles are also overplotted: black dotted line for the total EVE spectra, blue dashed lines for the free-free continuum, and dot-dashed purple line for the \ion{He}{2} continuum.}
		\label{eve_convol}
	\end{center}
\end{figure*}

The effective areas for each of the seven EUV channels on AIA were retrieved using the {\sc aia\_get\_response} function available in {\em SolarSoftWare IDL} \citep{free98}. The {\sc evenorm} keyword was applied to normalize the wavelength response functions to give a better agreement with full-disk EVE observations. Each channel has a fairly symmetric Gaussian response function centered on its primary wavelength (red curves in Figure~\ref{eve_convol}). 

\begin{figure*}[!t]
	\begin{center}
		\includegraphics[width=18.5cm]{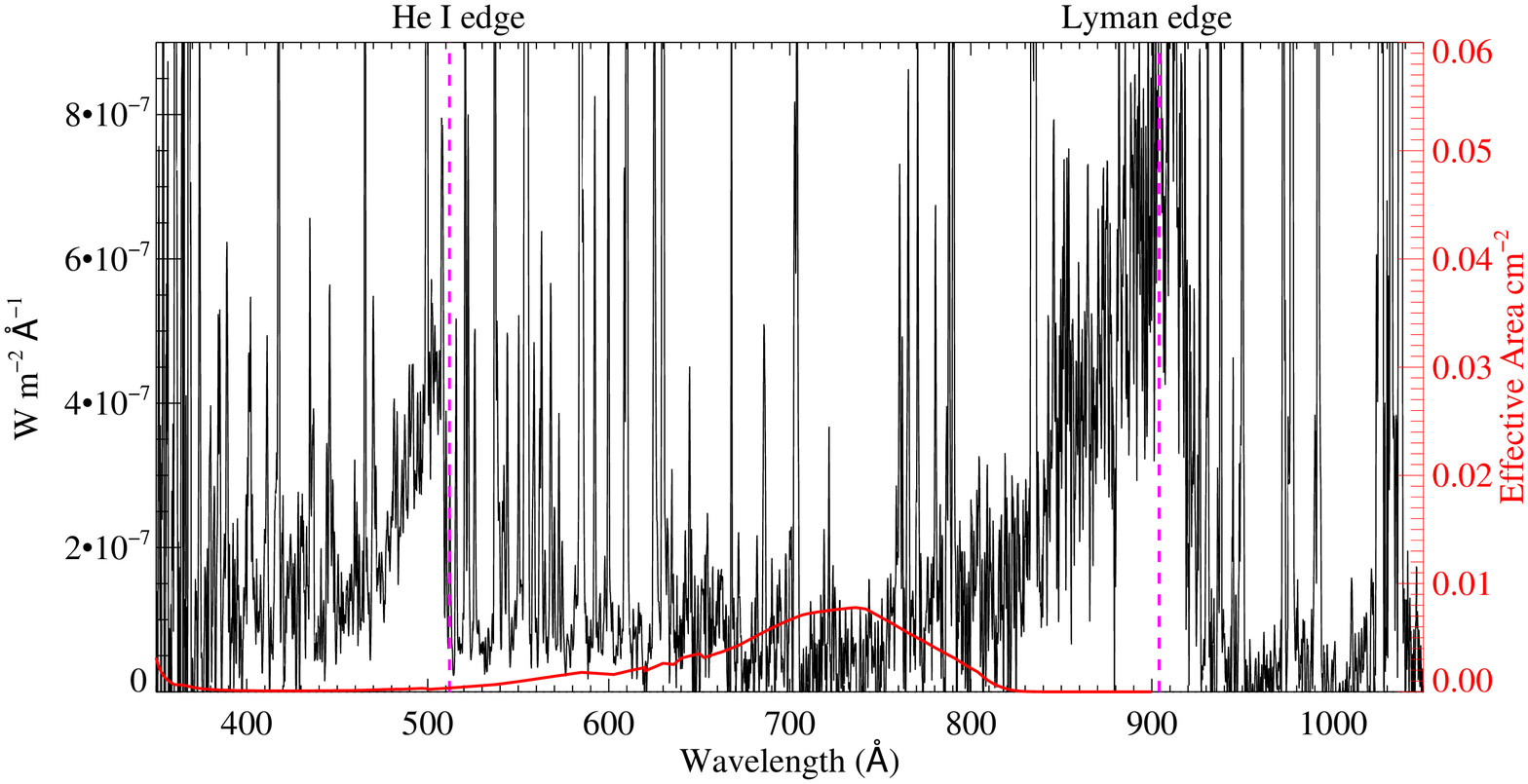}
		\caption{EVE MEGS-B spectrum, from 350--1050\AA, also taken at the peak of the 2011 February 15 flare (pre-flare subtracted), plotted using the same $y$-axis scaling in Figure~\ref{eve_convol}g. Overlaid is the component of the 335\AA\ response function (red curve) that spans this wavelength range. The ``red leak'' can be seen as a hump between 500--800\AA. The recombination edges of \ion{He}{1} and H Lyman are also plotted as vertical dashed purple lines.}
		\label{red_leak}
	\end{center}
\end{figure*}

However, the response function for the 335\AA\ channel also shows a considerable contribution at shorter wavelengths. \cite{boer12} predicted a second-order peak around 184\AA\  due to mirror reflectivity, which is shown in Figure~\ref{eve_convol}g. The existence of crosstalk at 131\AA\ in this channel was also predicted. Each telescope on the AIA instrument carries two channels which are grouped to minimise crosstalk. However, the effects are non-negligible for the wavelength response function of the 335\AA\ channel which shares telescope 1 with the 131\AA\ channel. Focal-plane filters select the required channel for data acquisition, but the 335\AA\ focal-plane filter also receives some light from the highly-reflective 131\AA\ channel. More recently, a significant ``red leak'' was discovered in the 335\AA\ channel during updated calibration measurements of the reflectivity of the AIA mirror coatings (Paul Boerner; Priv. Comm. See also \citealt{souf12}). This leak is seen as a bump in the 335\AA\ response function between 500--800\AA, which is covered by the EVE MEGS-B channel, as shown in Figure~\ref{red_leak}. This feature appears to lie between the \ion{He}{1} and H Lyman recombination continua. Neither of these continua is therefore expected to contribute significantly to the 335\AA\ channel during flares, as the response function at these wavelengths is also a factor of 6 weaker than those in the MEGS-A channel. However, the Lyman continuum may indeed contribute to the 335\AA\ channel during quiescent Sun conditions.

\section{Results}
\label{results}

\subsection{Spectral Convolution for the 2011 February 15 Flare}

Figure~\ref{eve_convol} shows the pre-flare subtracted spectra from EVE (black histogram) around the central wavelength range of each of the seven AIA channels at a time close to peak SXR emission of the 2011 February 15 flare. The black dotted lines show the EVE spectra having been convolved with the AIA effective areas and so gives the total predicted irradiance observed by each of the AIA passbands. The solid blue line shows the fit to the free-free continuum obtained using the method described in Section~\ref{data_anal}, while the blue dashed line shows this fit convolved with the AIA effective areas for each passband. The effective areas themselves are overplotted in red. In Figure~\ref{eve_convol}e, the fit to the \ion{He}{2} continuum around the 211\AA\ channel is overplotted in purple. The convolution of this fit with the response function is shown as a dashed purple line.

At the peak of the flare the 94\AA\ channel showed a strong line at 93.93\AA\ due to \ion{Fe}{18} above the underlying free-free emission (Figure~\ref{eve_convol}a), while the 131\AA\ channel appeared to be dominated by \ion{Fe}{21} and \ion{Fe}{23} at 128.75\AA\ and 132.91\AA, respectively (Figure~\ref{eve_convol}b). Results for the 171\AA\ channel were somewhat ambiguous as the \ion{Fe}{9} line at 171.07\AA\ that is believed to dominate this channel appeared to be very weak, or possibly even non-existent, as evidenced in Figure~\ref{eve_convol}c. This is possibly due to coronal dimming at these temperatures as noted by \cite{wood11} in several events observed by EVE, but could also be due to the fact that pre-existing material at \ion{Fe}{9} temperatures will have been heated to higher ionization states at this stage of the flare.

Emission in the 193\AA\ channel (Figure~\ref{eve_convol}d) appeared to be mostly due to \ion{Fe}{24} emission at 192.03\AA\ at this stage of the flare, with a weak contribution from \ion{Fe}{12} at 195.12\AA. The 211\AA\ channel (Figure~\ref{eve_convol}e) comprises multiple lines formed over a range of temperatures: \ion{Fe}{11} (209.78\AA), \ion{Fe}{13} (209.62\AA), \ion{Fe}{14} (211.32\AA), \ion{Fe}{17} (204.67\AA), and \ion{Ca}{16} (208.60\AA), all of which are considerably weak. This could also be due to coronal dimming at these temperatures, or heating of these elements to higher ionisation stages. Strong contributions from both free-free and free-bound (\ion{He}{2}) continuum emission are also present. The 304\AA\ channel is dominated by the self-blended \ion{He}{2} line at 303.78\AA\ (Figure~\ref{eve_convol}f) which is the strongest line in the EVE spectrum. Figure~\ref{eve_convol}g shows that the 335\AA\ channel comprises a strong contribution from \ion{Fe}{16} at 335.41\AA\ as well as contamination from the \ion{Fe}{21} and \ion{Fe}{23} lines, as well as continuum emission, around 131\AA\ as discussed in Section~\ref{data_anal}. However there did not appear to be any significant contribution from the second-order peak at 184\AA\ as predicted.	

\begin{figure*}
	\begin{center}
	\includegraphics[width=\textwidth]{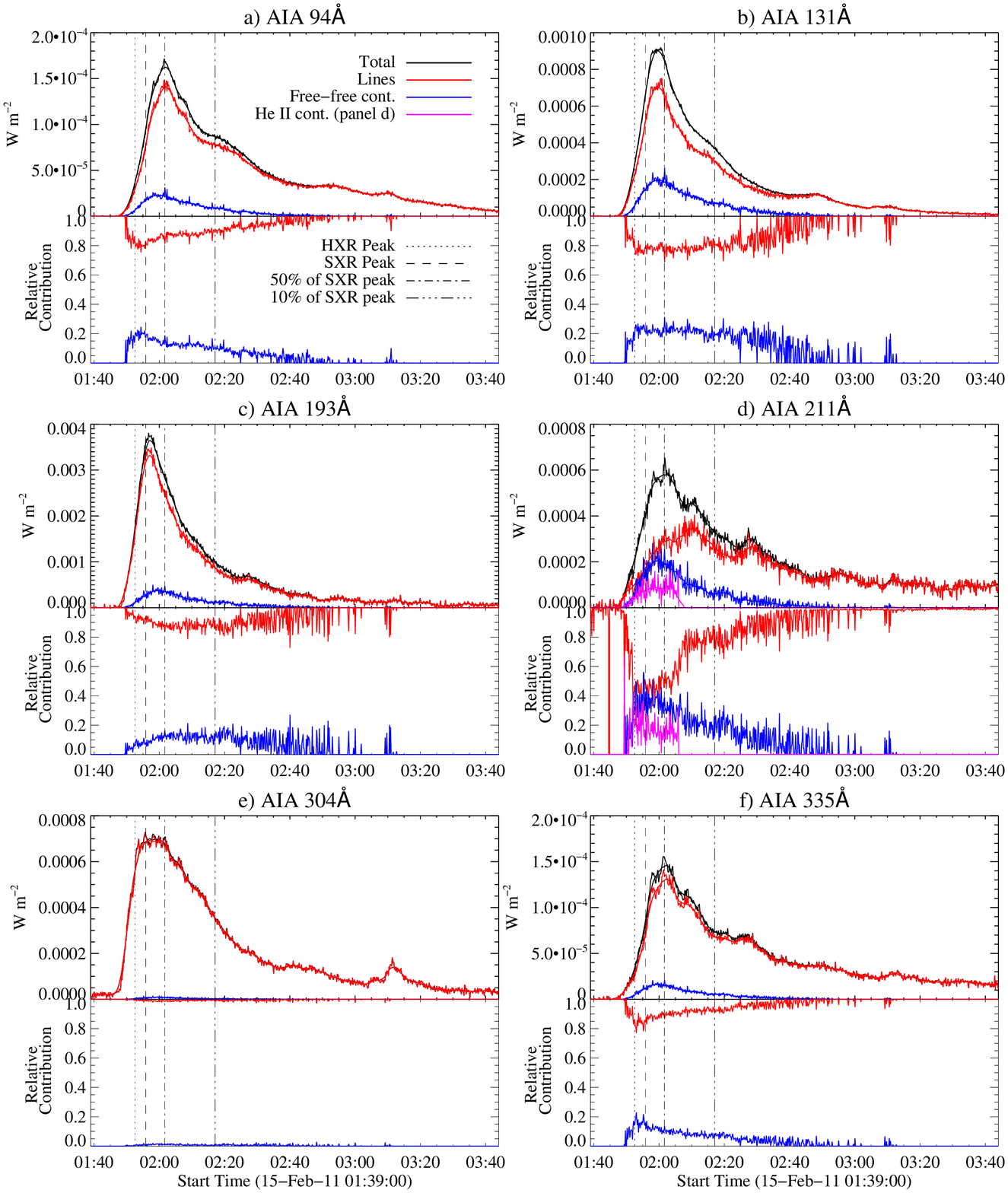}
	\caption{Lightcurves of the predicted emission in each of the AIA channels (except 171\AA) during the 2011 February 15 flare. Upper panels: Total EUV irradiance from EVE data convolved with AIA effective areas (black) for each of the seven AIA channels. Continuum contribution (blue) as measured from exponential fits convolved with AIA Effective area; lines emission (red) calculated from the difference between total and continuum emission. Lower panels: Relative contributions of continuum (blue) and line emission (red). The vertical solid, dotted, dashed, dot-dashed, triple-dot-dashed, and long-dashed lines denote the times of the flare onset, peak of the GOES derivative (a proxy for the HXR peak), the peak of the SXR emission, and points in the X-ray decay phase when the flux was 50\%, 10\%, and 1\% of the maximum, respectively.}
	\label{ff_ratios}
	\end{center}
\end{figure*}

\subsection{Continuum Time Profiles for the 2011 February 15 Flare}

Using the method described in Section~\ref{data_anal} lightcurves of the total emission predicted for each AIA channel were compiled for the duration of the flare (solid black curves in upper panels in Figure~\ref{ff_ratios}) between 01:40--03:40~UT for each passband. Lightcurves of the corresponding line (red) and continuum (blue) emission are also plotted. To draw more discernible comparisons, the relative fractions of each emission are shown in the lower panels of each plot. Each curve is also overlaid with a 6-point (1~minute) boxcar smoothed function from which the relative contribution at various stages during the flare were ascertained (Table~\ref{prct_cnt} and vertical lines in Figure~\ref{ff_ratios}).

The 94\AA\ channel might have been expected to have shown the strongest contribution of continuum emission given that free-free emission increases at shorter wavelengths. From the ratio plots in Figure~\ref{ff_ratios}a it can be seen that the continuum is particularly significant at the flare onset (15--20\%) but gradually decreases to about 10\% by the time the GOES flux reaches 10\% of its maximum before eventually returning to zero. The continuum also makes a significant contribution to the 131\AA\ channel. Throughout the course of the flare it makes up about 20--30\% of the total emission before decreasing during the late decay phase (Figure~\ref{ff_ratios}b). The continuum in the 193\AA\ channel increases from 0--15\% over the main phase of the flare before returning to zero (Figure~\ref{ff_ratios}c). 

Free-free continuum emission was found to be comparable to line emission in the 211\AA\ channel, making up between 40--50\% of the total emission around the peak of the flare. In Figure~\ref{ff_ratios}d the ratios shown for this channel are quite noisy, perhaps due to the existence of several lines, each being more prevalent at different stages of the flare, as well as scatter due to the fits to the continuum. \citealt{odwy10} also predicted the 211\AA\ channel to have the greatest proportion of continuum emission, but their estimate of 41\% was only applicable to the decay phase. The findings presented here suggest that only 25\% of emission in the decay phase is due to free-free emission during this event. Emission from the \ion{He}{2} continuum on the other hand, not considered by \cite{odwy10}, was found to peak around the time of peak HXR emission as expected by \cite{mill12}. At this time it made up $\sim$20\% of the total emission in the 211\AA\ channel after which it began to decrease. No reliable measurements of the \ion{He}{2} continuum could be made after about 02:05~UT.

The 304\AA\ channel predominantly observes emission from \ion{He}{2} at 303.78\AA. This is the strongest line in the EVE spectrum, both at quiescent and flaring times. Given that the continuum is inherently weaker at longer wavelengths, it is unsurprising that the continuum contribution to this channel is negligible ($<$1\%) in agreement with \cite{odwy10}.

Similar to the 94\AA\ and 131\AA\ channels, the lightcurves for the 335\AA\ channel (Figure~\ref{ff_ratios}f) show the greatest contribution from continuum emission at the flare onset ($\sim$20\%), which decreases to zero as the flare progresses. The continuum contribution to the 335\AA\ channel is primarily due to the crosstalk from 131\AA\ as the \ion{Fe}{16} line at 335.41\AA\ is very strong compared to the continuum at these wavelengths.

\begin{table}[!t]
	\begin{center}
		\caption{Relative Continuum Contribution to Each AIA Channel During the 2011 February 15 Flare}
		\label{prct_cnt}
		\begin{tabular}{lcccc} 
		\tableline
		\tableline
							& HXR		& SXR	& 50\% of		& 10\% of		\\
		Passband				& Peak		& Peak	& SXR Peak	& SXR Peak	\\
		\tableline
		94\AA  				& 0.19 		& 0.19 	& 0.13		& 0.11		\\
		131\AA 				& 0.21 		& 0.24 	& 0.23		& 0.22		\\
		171\AA 				& -  			& -  		& - 			& -  			\\
		193\AA 				& 0.08 		& 0.09 	& 0.13		& 0.15		\\
		211\AA\ (free-free)		& 0.35 		& 0.43 	& 0.37		& 0.25		\\
		211\AA\ (\ion{He}{2})	& 0.21 		&0.19 	& 0.14		& -			\\
		304\AA 				& 0.01 		& 0.01 	& 0.01		& 0.01		\\
		335\AA 				& 0.16		& 0.16	& 0.11		& 0.09		\\
		\tableline
	\end{tabular}
	\end{center}
\end{table}

Table~\ref{prct_cnt} summarizes the continuum contributions to each passband at four different stages of the flare: the hard X-ray (HXR) peak, the soft X-ray (SXR) peak, and during the decay phase when the SXR flux was at 50\% and 10\% of its maximum. The HXR peak was taken from the peak of the derivative of the GOES 1--8\AA\ light-curve (01:53:06~UT) under the assumption of the Neupert effect \citep{neup68}. The SXR peak was taken as the time of maximum X-ray emission (01:56:50~UT). These times are overplotted as vertical lines on each panel of Figure~\ref{ff_ratios} (dotted, dashed, dot-dashed, and triple-dot-dashed, respectively). 

\begin{figure}[!t]
	\begin{center}
		\includegraphics[width=0.5\textwidth]{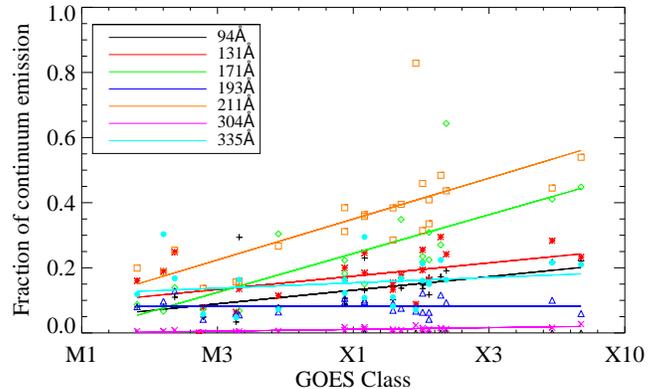}
		\caption{Plots of the continuum contribution to each of the AIA channels at the time of peak SXR emission as a function of GOES class. The tick marks on the $x$-axis denote steps of 0.1 dex in the 1--8\AA\ flux.}
		\label{goes_cont_frac}
	\end{center}
\end{figure}
	
\subsection{Continuum Contributions as a Function of GOES Classification}

The analysis performed on the 2011 February 15 X2.2 flare was repeated for 54 other flares of GOES class M2 or greater. Reliable fits to the free-free continuum were achieved in 24 events for all channels except the 171\AA. In this case the continuum level could only be measured in 18 flares. Figure~\ref{goes_cont_frac} shows how the free-free continuum contribution to each channel at the SXR peak of each event changes with respect to GOES class for flares greater than M2. For the 193\AA\ and 304\AA\ channels the continuum contribution does not change appreciably with respect to GOES class, with both channels exhibiting less than 10\% continuum emission. The 94\AA, 131\AA, and 335\AA\ channels show a slight dependence on X-ray classification but still contain less than 20\% continuum emission in all but the largest events. The 171\AA\ and 211\AA\ channels, however, both reveal a strong correlation between continuum contribution and peak X-ray flux. For the case of the 211\AA\ channel the continuum contribution increases from 20\% for M2 flares up to $\sim$55\% for the X6.9 flare of 2011 August 9. In one outlying event (an X1.9 flare that occurred on 2011 November 3) this increases to over 80\%. The distribution for the 171\AA\ channel shows a similar slope but has a maximum contribution of 45\% in the largest flares.

\section{Summary and Conclusions}
\label{conc}

Detailed measurements of the amount of continuum emission (both free-free and free-bound) contributing to each of the EUV passbands on SDO/AIA throughout an X-class solar flare are presented. Background-subtracted EVE spectra, and the corresponding fits to the underlying continua, were convolved with the AIA response functions to determine the relative amount of line and continuum emission in each channel as a function of time throughout the flare. The findings are in broad agreement with those of \cite{odwy10} who performed a similar analysis using synthetic spectra from CHIANTI. Both studies show that the 211\AA\ channel exhibits the largest contribution of free-free continuum emission during flaring conditions. The results presented show that this continuum can contribute between 40--50\% to the total emission in the 211\AA\ channel at the soft X-ray peak of an X-class flare. It is also shown that the recombination continuum of \ion{He}{2} also made up $\sim$20\% of the 211\AA\ channel emission. This free-bound emission was not considered in the work of \cite{odwy10}. While continuum emission in the \ion{He}{2} 304\AA\ channel was indeed negligible for the duration of the flare, and measurements for the 171\AA\ channel were uncertain due to potential coronal dimming at these wavelengths, typically between 10--25\% of emission observed in the remaining EUV channels was found to be due to free-free bremsstrahlung for the 2011 February 15 flare. 

This analysis was repeated for a number of flares between M2 and X7 and the level of continuum emission in both the 171\AA\ and 211\AA\ channels at the peak of each event was found to scale with respect to its GOES classification. This is due to the primary lines in each of these channels (\ion{Fe}{9} and \ion{Fe}{14}, respectively) becoming inherently weaker during flares, due to either coronal dimming or depletion of material at these temperatures as it gets heated to higher ionizations states, and therefore the relative intensity of the underlying continuum increases. 

EVE observations also help remove, or at least constrain, some of the assumptions used in generating synthetic flare spectra. For example, \cite{mill12b} identified three pairs of \ion{Fe}{21} ratios within the EVE spectral range which can be used to measure electron densities, particularly in large events. They found densities on the order of 10$^{12}$~cm$^{-3}$ for X-class flares; an order of magnitude larger than that assumed by \cite{odwy10} in their study. Similarly, \cite{warr12} have developed techniques to generate DEM profiles throughout a flaring event also using EVE data which would mark a significant advancement over the currently assumed flare DEM from \cite{dere79}. Several authors have also attempting to generate flare DEMs directly from AIA data (e.g., \citealt{hann12}, \citealt{batt12}, and \citealt{flet13}), but the resulting profiles may be inaccurate due to the underestimation or exclusion of the continuum emission in each of the filters. So while AIA often saturates during even moderate flaring events (e.g., \citealt{raft11}), the findings presented here underscore the need for coordinated spectral and imaging data to correctly interpret the observations. 

\acknowledgments
\noindent
The authors would like to thank Prof. Mihalis Mathioudakis for useful and insightful discussions on this work, as well as financial support and constructive feedback on the manuscript, and to Dr. Dean Pesnell for informing us of the updated response functions. The anonymous referee also provided very constructive feedback which greatly improved the scope of this paper. ROM is grateful to the Leverhulme Trust for financial support from grant F/00203/X, and to NASA for LWS/TR\&T grant NNX11AQ53G.

\end{document}